\documentclass[sigconf]{acmart}

\usepackage{booktabs} 

\usepackage{hyperref}
\usepackage{graphicx}
\usepackage{dirtytalk}
\usepackage{array, makecell}
\usepackage{amsfonts}
\usepackage{amsmath}
\usepackage{listings}
\usepackage{multirow}
\usepackage{color}
\usepackage{xcolor}
\usepackage{subfig}
\usepackage{diagbox}
\usepackage{standards}

\newcommand{\eg}[0]{e.g.\ }

\DeclareMathOperator*{\argmax}{arg\,max}

\newcommand{\pf}{PubMed Full}
\newcommand{\ptone}{PubMed T1}

\newcommand{\econ}{EconBiz}
\newcommand{\ef}{\econ~Full}

\newcommand{\eteight}{\econ~T8}

\newcolumntype{?}{!{\vrule width 1pt}}

\definecolor{info}{HTML}{1E90FF} 

\copyrightyear{2018} 
\acmYear{2018} 
\setcopyright{none}
\acmConference[JCDL '18]{The 18th ACM/IEEE Joint Conference on Digital Libraries}{June 3--7, 2018}{Fort Worth, TX, USA}
\acmBooktitle{JCDL '18: The 18th ACM/IEEE Joint Conference on Digital Libraries, June 3--7, 2018, Fort Worth, TX, USA}
\acmPrice{15.00}
\acmDOI{10.1145/3197026.3197039}
\acmISBN{978-1-4503-5178-2/18/06}
\renewcommand\footnotetextcopyrightpermission[1]{}

\begin{document}


\title{Using Deep Learning for
Title-Based Semantic Subject Indexing to Reach Competitive Performance to
Full-Text}

\author{Florian Mai}
\affiliation{%
  \institution{Kiel University}
}
\email{stu96542@informatik.uni-kiel.de}

\author{Lukas Galke}
\affiliation{%
  \institution{Kiel University}
}
\email{lga@informatik.uni-kiel.de}

\author{Ansgar Scherp}
\affiliation{%
  \institution{Kiel University}
}
\email{asc@informatik.uni-kiel.de}

\begin{abstract}
For (semi-)automated subject indexing systems in digital libraries, it is often
more practical to use metadata such as the title of a publication instead of the
full-text or the abstract.
Therefore, it is desirable to have good text mining and text classification algorithms that operate well already on the title of a publication.
So far, the classification performance on titles is not competitive with the
performance on the full-texts if the same number of training samples is used for training. 
However, it is much easier to obtain title data in large quantities and to use
it for training than full-text data.
In this paper, we investigate the question how models
obtained from training on increasing amounts of title training data compare to
models from training on a constant number of full-texts.
We evaluate this question on a large-scale dataset from the medical domain
(PubMed) and from economics (\econ).
In these datasets, the titles and annotations of millions
of publications are available, and they outnumber the available full-texts by
a factor of 20 and 15, respectively. To exploit these large amounts of data to
their full potential, we develop three strong deep learning classifiers and
evaluate their performance on the two datasets.
The results are promising. On the \econ~ dataset, all three classifiers
outperform their full-text counterparts by a large margin.
The best title-based classifier outperforms the best full-text method by 9.4\%. 
On the PubMed dataset, the best title-based method almost reaches the
performance of the best full-text classifier, with a difference of only 2.9\%.

\end{abstract}

%
%



\maketitle

\section{Introduction}\label{sec:introduction}
Semantic annotations are crucial for users of digital
libraries as they enhance the search of scientific documents. Given
the large amount of new publications,
automatic annotation systems are a useful tool for human expert annotators working at digital libraries
to classify the publications into categories from a (hierarchical) thesaurus.
However, providing automated recommendations for subject indexing in such
systems is a challenging task. This is partly due to the data from which
recommendations may be generated. Often neither the full-text of a publication
nor its abstract may be available.
For instance, the digital library
EconBiz contains only for 15\% of the documents an abstract. Even when
the content can be legally provided by the library to the end users, copyright
laws or regulations of the publishers may prevent text mining. Moreover,
collecting and processing PDFs where it is possible, e.\,g., for some Open
Access documents, adds high computational requirements to the library.
This puts annotation methods on demand that are based on data with better
availability, such as the title. Previous work by Galke et
al.~\cite{galke2017multilabel}, however, has shown that title-based methods
considerably fall behind full-text methods in terms of performance when the
number of samples for training is equal. If our classifier was a human expert,
this would not be a surprising result. A full-text contains more information and
therefore also more indication of the publication's topic. A human expert will
always make better annotations based on the full-text. In fact, the annotations
that are used as gold-standard for automated subject indexing experiments are
often created based on the full-text.

However, we argue that machine learning algorithms work differently than a
human. In contrast to a human, they often require hundreds of thousands or even
millions of training data to yield satisfactory models~\cite{brain1999effect}.
These amounts of data are not always available in the real world. One common reason is that
human expertise is required for creating a large enough gold standard, which is
expensive. For semantic subject indexing, the availability issues
mentioned above do not only come into play at prediction time, i.e.,
when a machine learning model is used in a productive system, but also during training. In effect, methods
based on the full-texts have drastically less training data available than
methods based on titles. This raises the question if title-based methods can
potentially narrow the performance gap to full-text methods by fully incorporating all training
data available. 

In this paper, we address this question. Formally, semantic indexing is framed
as a multi-label classification problem, where a (commonly small)
subset of labels has to be selected from a (relatively large) set of labels. From two digital libraries of
scientific literature,
PubMed and
EconBiz, we have compiled an English
full-text dataset and an English title dataset. Our compiled
datasets are quite different with respect to their
size.
From PubMed, we extracted 12.83 million titles. For 5\% of these a full-text is
available (646k).
From EconBiz, we extracted 1.06 million titles, of which approximately 7\% have a full-text (71k). In order to fully utilize these large amounts of data, we develop and
compare three different classifiers that have emerged from the deep learning
community in recent years. Deep learning has advanced the state-of-the-art in
many fields, such as vision, speech, and text~\cite{lecun2015deep}. These techniques are
known to shine when a lot of training data is available. For text
classification in particular, recent work~\cite{zhang2015character} suggests
that deep learning starts to outperform strong traditional models when 650k or more training samples are available.
The number of full-text in PubMed is right at the edge of this
number, whereas EconBiz has far less full-texts, making these datasets an
interesting and revealing choice.
In natural language processing, different types of neural networks have been
successfully employed on different tasks, but it is an open question whether
convolutional neural networks (CNNs), recurrent neural networks (RNNs), or multi-layer-perceptrons
(MLPs) are superior for text classification tasks.
Therefore, we employ a representative of each type in our study. 
We compare them against another strong MLP baseline (\emph{Base-MLP}), which has
previously been shown to also outperform traditional bag-of-words classifiers such as SVMs,
Naive Bayes, and kNN~\cite{galke2017multilabel}.
Since the label space in our datasets is very large, our study can be understood
as \emph{eXtreme Multi-Label Classification} (XMLC). Here, only few studies have
leveraged deep learning techniques to tackle the considerably harder problem
when the label space is large~\cite{zhang2017deep, liu2017deep}.
Hence, with our study, we contribute to the knowledge in this field, as well.

The results of our study indicate that title-based methods can match or even
outperform the full-text performance when enough training data is available.
On \econ, the best title classifier (MLP)
performs on par with the best full-text classifier (MLP) when training with only
8$\times$ as many titles as there are full-texts available.
When all available titles are used (approximately 15$\times$ more than full-texts), the title-based MLP
outperforms its full-text counterpart by 9.4\%.
On the PubMed dataset, the best title method is the RNN, and it almost reaches
the best full-text performance produced by an MLP.
The gap is only as small as 2.9\%. 
When using the same number of titles as full-texts are available, the gap in
classification performance is 10.7\%, indicating a considerable benefit from leveraging all title data.
Generally, the MLP performs well, outperforming RNN and CNN in three out of the
four combinations of the two datasets and title/full-text. It also consistently
outperforms the baseline when all titles or full-texts are used. Moreover, our
analysis suggests that our proposed classifiers are well-chosen for our study,
because they benefit from increasing amounts of training data better than the baseline.
The RNN performs rather poor on full-text, but shows strong performance on
titles.
While also yielding reasonably good results with CNN, it performs clearly below the other classifiers in all cases. 
This is a surprise, because a lot of the recent literature on large-scale text classification has
focused on CNNs (\eg~\cite{zhang2015character,conneau2016very,le2017convolutional}, also see
Section~\ref{sec:main-paper:relevant-work}).
Thus, our results indicate that it may be worth it to shift the research focus more towards other types of neural networks for text classification.

Our contributions can be summarized as follows:
\begin{itemize}
  \item For the first time, we study the question whether
  title-based methods can reach the performance of full-text-based methods by
  exploiting the surplus of available training data.
  \item We demonstrate that title-based methods are on par or even
  outperform full-text methods when the number of training samples is
  sufficiently large.
  \item We develop and compare three strong classifiers for (extreme)
  multi-label text classification, contributing to the debate on which type
  of neural network is superior for text classification.
\end{itemize}

The remainder of the paper is structured as follows. In the subsequent section, we
review relevant work on the comparison of short texts and full-text and
on deep learning for text classification.
In Section~\ref{sec:main-paper:methods}, we describe our deep learning models. We introduce the datasets and experimental
setup in Section~\ref{sec:main-paper:experimental-setup}, and present the
results in Section~\ref{sec:main-paper:results}. We analyze and discuss the
results in Section~\ref{sec:main-paper:discussion}, before we conclude.

\section{Related Work}\label{sec:main-paper:relevant-work}
In this section, we review previous literature relevant to our study. 
First, we discuss papers that compare performance on titles with performance on full-text.
Next, we briefly discuss methods for
multi-label text classification other than deep learning.
Finally, we discuss current deep learning methods for text classification.

\paragraph{Title versus Full-Text}
The work directly related to our study is the one by Galke et
al.~\cite{galke2017multilabel}. The authors compare titles with full-texts
for multi-label text classification on four datasets. 
Two datasets consist of scientific publications and are therefore comparable to the datasets in this study. 
In their experiments, the authors used the same number of samples for the
title-based methods and the full-text methods.
They found that the title-based methods can yield reasonably good performance.
However, the difference between title and full-text on the two scientific
datasets is still 10\% and 20\% in favor of the full-text, respectively.
The first dataset is from the economics domain, and it is an earlier version of the one used in this study.
The second dataset is from the political sciences. 
In their comparison of classifiers, an MLP to which in this study we refer as
\emph{Base-MLP}, outperforms all other (non-neural) classifiers in 7 out of 8
combinations. All the presented classifiers are based on the bag-of-words (BoW)
feature representation, a traditionally strong baseline for text classification that disregards word order. Due to clearly superior performance, Base-MLP can also
be considered the best representative of traditional BoW models.
We therefore report its performance as our baseline for all subsequently
developed models.

The comparison of metadata vs. full-texts has also been studied for
tasks other than document classification. Nascimento et
al.~\cite{nascimento2011source} studied how well queries for web search can be generated from title, abstract, or body, respectively. 
These queries are then issued to Web information sources to retrieve papers for recommendation. 
In their experiments, abstracts yielded the best queries, closely followed by the body. 
The title clearly performs worse. 
On the contrary, Nishioka and Scherp~\cite{nishioka2016profiling} have
demonstrated that competitive paper recommendations to researchers based on
their Twitter profile can be made by using only the title of the paper. This is
achieved by their novel profiling method HCF-IDF, which is able to extract sufficient conceptual
information from the title through spreading activation over a hierarchical
knowledge base. Galke et al.~\cite{galke2017word} use text embedding techniques
for the information retrieval task and compare how well these techniques
perform when the index is built upon the title, abstract, or full-text of the documents.
Here, titles have demonstrated a clear advantage over abstract and full-text. 
Lastly, Hemminger et
al.~\cite{hemminger2007comparison} compare full-text search with metadata
search in the PubMed database, where full-text search yields better results.

\paragraph{Multi-label Text Classification}
Text classification is a well-studied problem.
k-Nearest-Neighbors and SVM are common choices for text classification
(see \eg~\cite{joachims1998text, rubin2012statistical, grosse2015comparison}).
However, kNN's complexity grows in the number of training samples,
which is problematic for the training sample sizes we consider in our study.
SVMs do not provide a natural adaptation for multi-label classification. A
binary relevance classification scheme, however, is impractical when the number
of labels is large.

An active field of multi-label text classification research is the MeSH
indexing community. This area is concerned with annotating PubMed articles with medical
subject headings. BioASQ~\cite{bioasq} is a challenge that recently finished
its 5th iteration. It has the goal to advance the state-of-the-art in MeSH
indexing, and to this end provides large datasets for training. We
acknowledge the significance of the MeSH indexing community, and in our study,
we include the dataset from the latest iteration of the BioASQ challenge.
However, many of the approaches successful at this challenge, including learning
to rank~\cite{huang2011recommending, liu2015meshlabeler, mao2017mesh} and
pattern matching~\cite{mork2014recent}, make use of features tailored to the
biomedical domain. 
Since in this paper, we study domain-independent methods for subject indexing
in digital libraries, these methods are not appropriate.

\paragraph{Deep Learning for Text Classification}

Some early works leverage neural networks for multi-label text classification.
In order to capture label inter-dependencies, Zhang and
Zhou~\cite{zhang2006multilabel} employ a pairwise ranking loss function for text
classification. Nam et al.~\cite{nam2014large} show that replacing 
the ranking-loss with cross-entropy leads to faster convergence and overall
better prediction performance. Furthermore, they are the first to incorporate
some of the milestone advancements from the deep learning era like rectified linear units, dropout, and the smart optimizer AdaGrad.

In recent years, neural networks have become the state-of-the-art in multi-class
text classification, outperforming traditional linear BoW models, in
particular on very large datasets.
On a diverse set of text classification datasets of rather small
scale (up to 10.7k samples), neural networks have shown their capability to
perform well. These include CNNs~\cite{kim2014convolutional,
yin2016multichannel} and RNNs~\cite{tai2015improved}, as well as a combination
of both~\cite{zhang2016dependency}.

Zhang et al.~\cite{zhang2015character} were the first to introduce several
large-scale multi-class text classification datasets ranging from 120k to
3.6 million training samples. The number of classes ranges from two to
14.
Zhang et al. proposed a deep, character-based convolutional neural network and
compared it with a number of traditional models, including multinomial logistic
regression based on bag-of-ngrams with TF-IDF, and deep learning models such as
a Long Short-Term Memory network (LSTM) and word-based CNNs.
The finding most relevant to our study is that traditional models tend to
outperform the neural network architectures on the four relatively small
datasets with 560k training samples or less, whereas on the remaining four
datasets with 650k training samples or more their neural network approach is
superior. Although these numbers may vary depending on the dataset and
classification task at hand, this result is the main reason why we choose to
employ deep learning techniques in our study.

Several studies employing deep learning on these datasets have followed. Conneau
et al.~\cite{conneau2016very} draw inspiration from the computer vision
community and improve the performance of character-based CNNs by increasing the
depth. Le et al.~\cite{le2017convolutional} put these results into
perspective by demonstrating that a shallow, word-based CNN performs
on par with these models or better. Therefore, we have limited our study to
shallow, word-based CNNs. For text classification, it is an open question
whether CNNs or RNNs are superior~\cite{yin2017comparative}. Consequently,
LSTMs had also some success on these large-scale
datasets~\cite{yang2016hierarchical, yogatama2017generative}, as well as hybrid
approaches~\cite{xiao2016efficient}. Joulin et
al.~\cite{joulin2016bag} demonstrated that even a linear MLP classifier can
yield results competitive with non-linear deep learning methods while
maintaining computational efficiency.

To the best of our knowledge, each of the neural network types MLP, CNN, and
LSTM provide the current state-of-the-art on at least one of these large-scale datasets (as
shown in the work by Le et al.~\cite{le2017convolutional}). Hence, with our study we
would like to contribute to this discussion by employing a representative of
each of the neural network types.

Finally, due to the number of labels in our datasets, our study may be
classified as XMLC. To the best of our knowledge,
only two papers studied the application of deep learning to this
setting~\cite{zhang2017deep, liu2017deep}, from which we draw some inspiration
for our models, which are described below.

\section{Methods}\label{sec:main-paper:methods}
In this section, we present three neural network architectures for text
classification used in our study. In the design phase, we aimed at carefully developing the
strongest representative of each of the most common types of neural networks,
MLPs, CNNs, and RNNs\footnote{For transparency, intermediate
results of the design phase can be found in an extended version of the paper
provided at \url{https://github.com/florianmai/Quadflor}.}. Here, we only present the final
architectures which are used in the experiments presented in
Section~\ref{sec:main-paper:experimental-setup}. These architectures may differ
depending on the dataset and type of text (title or full-text) they operate on.
This is necessary to avoid a bias for either of the datasets or types of
text.
The output layer as well as the training procedure in our neural networks are
the same for all neural network architectures presented here. 

Below, we describe the training procedure, before we go
into each neural network architecture up to the last hidden layer. For brevity,
we omit a formal mathematical description of the models here.

\subsection{Training Procedure}\label{sec:main-paper:training-procedure} The semantic
annotation task is formally a \emph{multi-labeling problem}, where instead of belonging to exactly one class, each publication is assigned a set of
labels. This is an important difference to most of the previous literature on
text classification with large datasets. \emph{Binary Relevance} is a
common technique to adapt a classifier for a multi-labeling problem. However,
this is a costly technique when the number of labels is high because it requires
to train as many classifiers as there are labels. Neural networks have a more
natural way to deal with multi-label classification, which is also made use of
by Galke et al.~\cite{galke2017multilabel}. For multi-class text
classification, the softmax activation function is used at the
output layer to obtain a probability distribution over the classes. For
multi-label classification, however, the sigmoid activation can be employed to determine a
probability $p_l$ for each label $l$ whether it should be assigned or not. The
difference is that softmax regards all labels at once, while sigmoid makes
an independent decision for each label. Finally, the binary decision whether
label $l$ is assigned is made by checking whether $p_l$ exceeds a threshold
$\theta$.

Using Adam~\cite{kingma2014adam}, the networks are trained as to minimize the sum of all
binary cross-entropy losses over all labels. This has shown to be superior to a
ranking-based loss on multi-label text classification~\cite{nam2014large,
liu2017deep}.
Training is executed in mini-batches of size 256. We employ early stopping
for regularization and as criterion to terminate training.
The performance on the validation set is evaluated in terms of the
sample-based $F_1$-measure. Training is terminated when the validation score has
not improved over the best reported score for 10 consecutive evaluations.

Since the output of the sigmoid activation function can be interpreted as the
probability whether a label should be assigned, a typical choice for the
threshold is $\theta = 0.5$. However, depending on the evaluation metric,
dataset, or model that generates the assignment probabilities, this value
does not necessarily yield optimal results.
Unfortunately, finding a good value for $\theta$ can be
computationally expensive, especially when the datasets are very large.
Therefore, we use a heuristic that continuously adjusts the threshold
\emph{during} training. To this end, the evaluation on the validation set used
for early stopping is also used to optimize $\theta$.

Formally, we initially set $\theta_0 := 0.2$, which Galke et
al.~\cite{galke2017multilabel} found to be a better threshold value than 0.5.
After each validation step $i$, where the classifier predicts a probability for each of the $\left| L \right|$
labels and each of the $n$ samples in the validation set, accumulated in $P_i
\in (0,1)^{n \times |L|}$, we set
\begin{equation*}
\theta_i := \argmax\limits_{\tilde{\theta} \in \{-k * \alpha + \theta_{i - 1}, \ldots, k * \alpha + \theta_{i - 1}\}} F_1(P_i; \tilde{\theta})
\end{equation*}
where $\alpha > 0$ is the step size and $k$ controls the number of threshold
values to check. This heuristic is based on the observation that the optimal
choice for $\theta_i$ is in most cases in close proximity to the optimal choice
of the previous evaluation step, $\theta_{i - 1}$. Since computing the $F_1$
score can be costly, we set $k = 3$ and $\alpha = 0.01$ to trade off granularity with speed.

In our preliminary experiments, this way of optimizing the threshold
consistently yields good results, sometimes even better than when it is
optimized manually.
\subsection{Multi-Layer-Perceptron}
Our baseline is a multi-layer-perceptron (MLP) described by Galke et
al.~\cite{galke2017multilabel}. It has one hidden layer with 1,000 units and
rectifier activation and it takes a TF-IDF~\cite{salton1988term} bag-of-unigrams
as input.
The bag-of-unigrams only contains the 25,000 most common unigrams, which we determined to be sufficient
for the multi-labeling task. For regularization, dropout~\cite{srivastava2014dropout} is applied
after the hidden layer with a keep probability of 0.5. We will refer to this
baseline as \emph{Base-MLP}.

We extend the MLP from Galke et al.~\cite{galke2017multilabel} by incorporating
some techniques inspired from recent deep learning literature.
The MLP architecture introduced in this study can be viewed as an adaptation of
fastText~\cite{joulin2016bag} for multi-label classification. FastText is a
linear BoW model enhanced with a feature sharing component (a hidden
layer with identity activation) and local word order information (bigrams).
We adopt this model by adding the 25,000 most common bi-grams in addition to the
25,000 most common unigrams. However, we found that omitting the non-linearity
at the hidden layer rather hurts the classification performance considerably.
Therefore, we keep the rectifier as a non-linear activation at the hidden layer.

In the introduction, it was mentioned that deep neural networks excel when the
number of training samples is very large. This is because the representational
power of neural networks increases as the number of parameters increases. This
can be obtained by adding more layers or by adding more units to the existing
layers. Additionally, deeper networks may be able to learn
hierarchical representations of the input, as can be observed in the vision
domain~\cite{he2016deep}. However, deep networks are generally harder to train
due to the vanishing and exploding gradient problems. We apply
Batch Normalization~\cite{ioffe2015batch} to our deep MLPs to alleviate those.

In summary, MLP differs from Base-MLP in that it
incorporates bi-grams and uses multiple layers and Batch Normalization where
appropriate.

\subsection{Convolutional Neural Network}\label{sec:main-paper:methods:cnn}
We present a CNN architecture whose core was introduced by
Kim~\cite{kim2014convolutional} for sentence classification and has since been
repeatedly adopted and enhanced upon. We adopt and combine some of these
enhancements for our model.

The CNN operates on word embeddings, which are initialized with a pretrained
model but finetuned during training. As in Kim's model, our CNN applies a 1D-convolution by sliding
a window over the text in order to extract features at each position. These
outputs are then transformed by a non-linear activation function (the detector).
Commonly, the most salient position is selected by applying max-pooling after
the detector stage. However, Liu et al.~\cite{liu2017deep} instead split the
output of the convolution into $p$ nearly equal chunks, and perform max-pooling
on each chunk. Afterwards, the outputs of the pooling stages are concatenated.
For $p = 1$, this is identical to Kim's architecture.

Commonly, this process is repeated for multiple window sizes.
The outputs of these processes are then concatenated before passing them to the next layer.
For example, Kim's CNN uses window sizes 3, 4, and 5, while Liu et al. use 2, 4,
and 8. We experimentally determined that using 2, 3, 4, 5, and 8 yields to even
better results.

In Kim's model, the concatenated output of the pooling stages is directly
propagated to the output layer. Liu et al. argue, however, that it is better to
have an additional fully-connected layer with $n_b$ units, called the
\emph{bottleneck layer}, because it adds more representational power to the network
through the increased depth.

Considering the complexity of our datasets and the number of training
samples available, the question of increased capacity arises with CNNs. Similar
to MLPs, an increase in capacity can be achieved either through wider
convolutions (larger feature-map) or additional stacked layers of convolutions.
Since a recent study by Le et al.~\cite{le2017convolutional} has shown that
depth does not yield improvement over shallow nets, we only consider the former
approach.

\subsection{Recurrent Neural Network}
The Recurrent Neural Network (RNN) is a family of neural networks that was
specifically designed for sequential input data. By maintaining a hidden
state,  the network is able to keep track of previous inputs. However,
the vanilla RNN has difficulties keeping track of inputs that are far in the
past. The LSTM~\cite{hochreiter1997long} was designed to alleviate this by
explicitly modeling the control over whether the current hidden state shall be
forgotten, updated, or kept. For this study, initially we use an LSTM that has already achieved
good results for text classification in a study by Zhang et
al.~\cite{zhang2015character}. This LSTM is the ``vanilla'' version described
in \cite{greff2017lstm}. Our final model, however, incorporates two techniques
which have proven useful for NLP tasks in recent years, attention and
bidirectionality~\cite{young2017recent}.

Since any RNN produces an output at every time step, the outputs have to be
aggregated after processing the entire sequence, in order to pass a vector of
fixed size to the next layer. We experimented with choosing the last output,
computing the sum, computing the average, and computing a weighted average where
the weights are determined by an attention mechanism as used by Yang et
al.~\cite{yang2016hierarchical}. While the benefit over the other aggregation
methods is not large for titles, the attention mechanism consistently
performs best. On full-texts, on the other hand, the difference is considerable.
This is intuitive, because there is less need to focus on specific parts of the input if
the input is as short as in a title.

In the same fashion as Yang et al.~\cite{yang2016hierarchical}, we incorporate
bidirectionality into our LSTM by concatenating at each time step the output of
an LSTM that reads the input sequence from left to right, and an LSTM that reads
the sequence in reverse order. This is a common technique to make the model
aware of the entire sequence at every time step, and commonly boosts
performance in text applications.

Again, we made some effort to investigate an increase in the capacity of the
LSTM to account for the large number of training samples in our
datasets. In the past, both increasing the memory cell size (the width) and
stacking LSTMs on top of each other has been successful in some NLP tasks.
For text classification, this has not been the case. The results of our
experiments support this, where wider LSTMs are superior to stacked LSTMs, even
when variational dropout~\cite{gal2016theoretically} is used.

\section{Experimental Setup}\label{sec:main-paper:experimental-setup}
\begin{figure*}[!t]
\includegraphics[width=2\columnwidth]{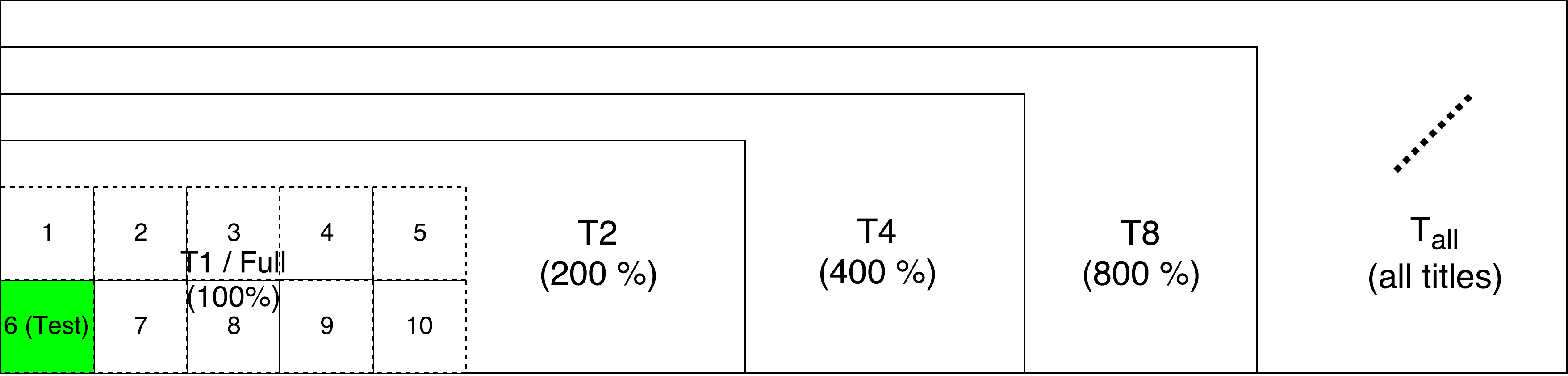}
\caption{We organize our dataset in several sub-datasets to perform an iterative
evaluation. T1 and Full comprise of the same set of publications (all samples where a full-text is available), and is split into
10 folds in order to perform the same 10-fold cross-validation with titles and
full-texts for a fair comparison. For T2, T4, T8,
increasingly more titles that do not have a full-text are added for training,
but are in each cross-validation step evaluated on the same test samples
(highlighted in green for exemplification). $T_{all}$ includes all title
samples.}
\label{fig:title-dataset-names}
\end{figure*}

\subsection{Datasets}\label{sec:dataset}

We built English datasets from two digital libraries of scientific publications.
EconBiz\footnote{\url{https://www.econbiz.de/}} is a search portal
for economics and business studies. Currently, it contains 2,485,000
English publications, out of which 615k are open access. Many
publications are annotated by experts with a variable number of subject headings
(so-called subject indexing) taken from a standardized set, the ``Thesaurus for
Economics'' (STW)\footnote{\url{http://zbw.eu/stw/version/latest/about}}.
From all 2,485,000 English publications, we filtered the ones that
have annotations and extracted their title.
After deleting duplicates,
1,064,634 publications with annotations remain.
Out of the 615k open access publications, the number of publications that have
annotations and whose full-text can be downloaded and processed reduces to
70,619, which is 6.63\% of all publications.

PubMed\footnote{\url{https://www.ncbi.nlm.nih.gov/pubmed/}} is a search engine
for biomedical and life science literature provided by the US National Library
of Medicine. The publications found on PubMed are annotated with ``Medical
Subject Headings'' (MeSH)\footnote{\url{https://www.nlm.nih.gov/mesh/meshhome.html}} by human
curators. We obtained a dataset consisting of millions of publication metadata,
including title and MeSH annotations, from the training set of the semantic indexing task
of the BioASQ challenge 2017~\cite{bioasq}, which are all in English language.
PubMed Central\footnote{\url{https://www.ncbi.nlm.nih.gov/pmc/}} is
an archive of full-texts of biomedical and life science literature provided by
the US National Library of Medicine. It comprises 4.3 million publications,
which can be accessed freely and which are mostly English. However, only 1.5
million are open access and therefore allow text mining. From this dataset, we
computed the intersection with the publications obtained from the BioASQ challenge. After
removing duplicates, 12,834,026 titles and 646,513 full-texts with respective
annotations remain. Hence, 5.04\% of the samples have a full-text.

Table~\ref{tab:datasets} lists some characteristics of the two datasets, EconBiz and PubMed. In
terms of combinatorial complexity, the PubMed dataset is a harder problem
because the number of labels out of which to pick the annotations for a
publication is much higher. Yet, due to the relatively large number of labels and
small number of samples per label on average, both datasets can be considered as
XMLC problems\footnote{An overview of datasets commonly considered XMLC can be
found at \url{http://manikvarma.org/downloads/XC/XMLRepository.html}.}.
The titles in PubMed contain on average more words than the publications'
titles in EconBiz. However, this fact is put into perspective considering that
the titles in PubMed  have on average more labels to be predicted than there are words in the title.
Regarding the full-texts, the ratio of words/labels is approximately the same in both datasets.
Another fact worth noting is that the titles corpora have on average one
label less than the full-texts. This suggests that the label
distributions in the title dataset and full-text dataset are quite different.
\begin{table}
\caption{Characteristics of \econ~and PubMed datasets. $\left| D \right|$
denotes the sample size, $\left| L \right|$ denotes the number of labels
used in the dataset, $d/l$ is the average number of publications a label is
assigned to. $l/d$ is the average number of labels assigned to a publication.
$\left| V \right|$ is the size of the vocabulary and $w/d$ denotes the average
number of words per document.}
\label{tab:datasets}
\resizebox{\columnwidth}{!}{
\begin{tabular}{|c||c|c|c|c|}
\hline
\multirow{2}{*}{} & \multicolumn{2}{c|}{\econ~ (STW)} &
\multicolumn{2}{c|}{PubMed (MeSH)} \\

& Title & Full-Text & Title & Full-Text \\
\hline
\hline
$\left| D \right|$ & 1,064,634 & 70,619 & 12,834,026 & 646,513 \\
\hline
Size & 78.8MB & 6.27GB & 1.32GB & 20.06GB \\
\hline
$\left| L \right|$ & 5661 & 4849 & 27773 & 26276 \\
\hline
$d/l$ & 819.1 & 75.8 & 5852.3 & 331.0 \\
\hline
$l/d$ & 4.4 & 5.3 & 12.6 & 13.5 \\
\hline
$\left| V \right|$ & 91,505 & 1,502,336 & 660,180 & 6,774,130 \\
\hline
$w/d$ & 6.88 & 6694.4  & 9.6 & 2533.4 \\
\hline
\end{tabular}} 
\end{table}

Please note that on both datasets, the set of titles is a superset of the
set of full-texts.

\subsection{Experiments}\label{sec:experiments}
In order to assess how the title-based
methods behave as more and more titles are considered for training, we create
sub-datasets of the title datasets by iteratively adding more
data. This is illustrated in Figure~\ref{fig:title-dataset-names}.

For a fair comparison of title and full-text performance, the trained model must
be evaluated on the same data. To this end, we split the set of publications
where a full-text is available into ten folds, and perform a 10-fold
cross-validation. In each iteration, nine folds are selected for training, and
one is selected for testing. From this training set, we randomly select 20\% for the
validation set for early stopping and adjusting the threshold, as described in
Section~\ref{sec:main-paper:training-procedure}.
These comprise the data used for our experiments on full-text, and will be
abbreviated as \ef~ and \pf, respectively.

For the experiments on titles, the same publications from the 10-fold cross-validation are used for testing.
However, the training set is iteratively extended with more samples from the
titles dataset, so that the total number of training samples is always a power
of two of the number of samples in the full-text experiment. In total, we conduct
experiments on five title sub-datasets per domain:
T1, T2, T4, T8, and $T_{all}$. Here, Tx means that $x$ times as many title samples are
used for training as there are full-text samples in the dataset.
Lastly, $T_{all}$ contains all title samples from the dataset.

We run each of the four classifiers Base-MLP, MLP, CNN, and LSTM on all of the sub-datasets. In total, we run
48 cross-validations.
Following Galke et al.~\cite{galke2017multilabel} who argue that the
sample-based $F_1$-metric best reflects how subject indexers work, we use this
metric to report the results. We also use this metric for early stopping and
threshold adjustment on the validation set.
The sample-based $F_1$-measure calculates the harmonic mean of precision and
recall for each sample individually, and averages these scores over all samples.

\subsection{Choice of Hyperparameters and Training}
Since there are a lot of tunable hyperparameters involved in deep learning,
tuning multiple hyperparameters at the same time can be very expensive,
especially when the datasets are very large. On the other hand, fixing
hyperparameters across all datasets and models would not be a fair approach in
our study because the datasets and model architectures are very different and
therefore may require very different hyperparameter settings.
As a compromise, we decided to tune the hyperparameters for full-texts and
titles separately in an incremental fashion on one fold. Here, we tuned one
hyperparameter at a time and selected the locally best solution for full-text and titles,
respectively. It is important to note that the
parameters for titles were determined based on the performance on $T_{all}$,
and were adopted for all other title sub-datasets. On the one hand, this
alleviates a lot of the computational cost and allows to compare the
performance between title sub-datasets. On the other hand, especially the
performance of smaller sub-datasets might be suboptimal due to overfitting.
This must be kept in mind for analysis.

In our experiments involving the MLP, we use a one-layer MLP
with 2,000 units and dropout with a keep probability of 0.5 after the hidden
layer for all experiments. Only for the experiments on the PubMed titles, we use a two-layer MLP with
1,000 units each, and apply no dropout. Instead, we use Batch Normalization after
each hidden layer. In all cases, the initial learning rate for Adam is set to
0.001.

In the CNN experiments, we use $p = 3$ chunks and $n_b = 1,000$ units at the
bottleneck layer~\cite{liu2017deep} for both full-text experiments. On titles,
we do not perform chunking ($p = 1$), and use a bottleneck layer size of $n_b =
500$. The size of the feature map is set to 400 in all experiments except
for \pf, where we use 100. The keep probability is set to 0.75 in all cases, and the initial learning rate is 0.001.

We use a single-layer LSTM for all experiments. For both datasets, we
determined 1,536 to be the best size for the memory cell when using titles. 1,024
units and 512 are used for \pf~and \ef, respectively. The keep probability is
set to 0.75 in all experiments except for PubMed on titles, where we set it to 0.5.
The initial learning rate is 0.01 for \ef~and 0.001 in all other cases. Training
is done with backpropagation through time by unrolling the LSTM until the end of the
sequence.

We adopt the preprocessing and tokenization procedure of Galke et.
al~\cite{galke2017multilabel}. For the LSTM and CNN, we use
300-dimensional pretrained word embeddings obtained from training
GloVe~\cite{pennington2014glove} on Common Crawl with 840 billion
tokens\footnote{This pretrained model can be downloaded at \url{https://nlp.stanford.edu/projects/glove/}.}. Out-of-vocabulary words are discarded.
The maximum sequence length is limited to the first 250 words. Longer sequences
were harmful in preliminary experiments.

For implementation of our neural network models, we used
the deep learning library TensorFlow\footnote{\url{https://www.tensorflow.org/}}
and integrated them within the multi-label classification framework
``Quadflor"\footnote{To increase the reproducibility of our study, we made the
source code and the title datasets available at \url{https://github.com/florianmai/Quadflor}.}.
All experiments are run either on an NVIDIA TITAN or on a TITAN Xp GPU which both have 12GB of RAM.

\section{Results}\label{sec:main-paper:results}
\begin{figure*}
\subfloat[]{\includegraphics[width=\columnwidth]{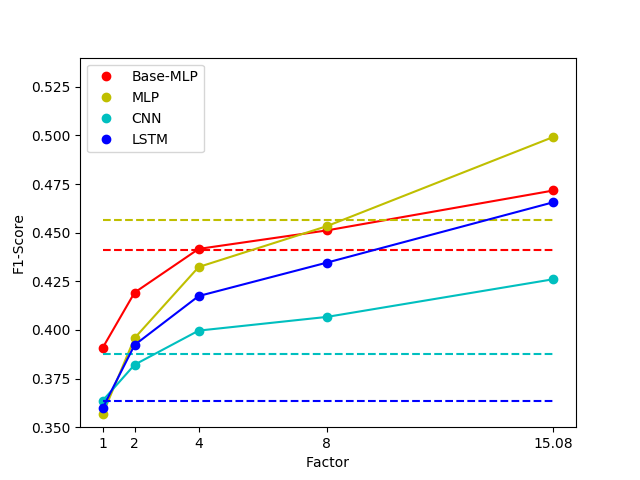}}
\subfloat[]{\includegraphics[width=\columnwidth]{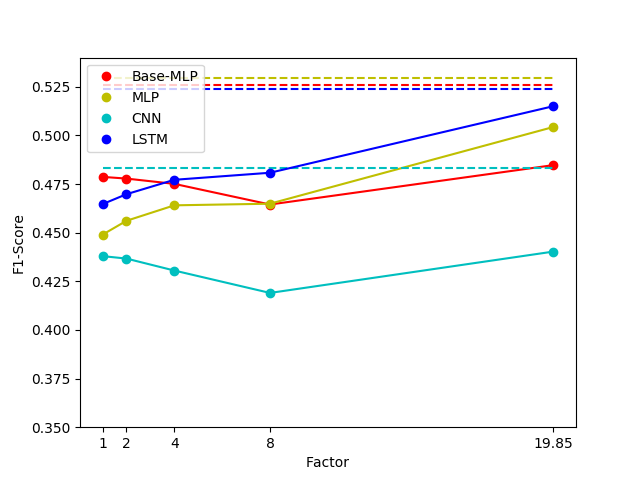}}
\caption{The figures show the performance of each classifier on titles as a
function of the sample size relative to the number of full-texts as a solid
line on \econ~ (a) and PubMed (b). The dashed horizontal lines represent the
respective classifier's performance on the full-text.}
\label{fig:results-plot}
\end{figure*}

The results of our experiments are shown in Table~\ref{tab:results}. In
addition, we plot the performance of each method as a function of the number of
samples used for training the title model. These are shown in
Figure~\ref{fig:results-plot}.
\begin{table*}
\caption{Results of experiments in terms of sample-based $F_1$-measure. The
best performing method on each sub-dataset is printed in bold font.}
\label{tab:results}
\resizebox{\textwidth}{!}{
\begin{tabular}{|c||c?c|c|c|c|c||c?c|c|c|c|c|}
\hline
  \multirow{2}{*}{\backslashbox{Method}{Dataset}} &
  \multicolumn{6}{c||}{\econ~ $F_1$ scores} &
  \multicolumn{6}{c|}{PubMed $F_1$ scores} \\
  \cline{2-13}
   & Full-Text & T1 & T2 & T4 &
  T8 & $T_{all}$ & Full-Text & T1 & T2 & T4 &
  T8 & $T_{all}$ \\
  \hline
  
  Base-MLP & 0.441 & \textbf{0.391} & \textbf{0.419} & \textbf{0.442} &
  0.451 & 0.472 & 0.526 & \textbf{0.479} & \textbf{0.478} &
  0.475 & 0.465 & 0.485 \\
  \hline
  MLP & \textbf{0.457} & 0.357 & 0.396 & 0.432 &
  \textbf{0.453} & \textbf{0.500} & \textbf{0.530} & 0.449 & 0.456 &
  0.464 & 0.465 & 0.504 \\
  \hline
  CNN & 0.387 & 0.364 & 0.382 & 0.400 &
  0.407 & 0.426 & 0.483 & 0.438 & 0.437 &
  0.431 & 0.419 & 0.440 \\
  \hline
  LSTM & 0.363 & 0.360 & 0.392 & 0.417 &
  0.435 & 0.466 & 0.524 & 0.465 & 0.470 &
  \textbf{0.477} & \textbf{0.481} & \textbf{0.515} \\
  \hline
\end{tabular}}
\end{table*}

\paragraph{\econ}
On the \econ~dataset, the best results on both titles and full-texts are
obtained by MLP. The title-based method is on par with the full-text
method when eight times as many titles as full-texts are used. When all titles
are used, the title-MLP outperforms its full-text counterpart by 9.4\%,
achieving an $F_1$-score of 0.500. In contrast, when the same number of samples
for full-text and titles is used, the gap between the best title method (Base-MLP) and the best
full-text method (MLP) is 16.9\% in favor of the full-text. 

The MLP seems to benefit the most from additional titles. Initially, when as
many titles as full-texts are used for training, all our proposed methods,
i.\,e., MLP, CNN, and LSTM, perform within 0.007 points in $F_1$-score from
each other, but MLP performs worst. However, the relation flips as more titles
are added. In Figure~\ref{fig:results-plot} (a), we can observe that MLP has the steepest curve of improvement out of all methods, in particular when considering the
improvement from T1 to T2. With twice as many titles as full-texts, MLP is
already the best performing classifier out of the ones we have proposed. The
gap to the other methods only gets wider as more data is added for training.
Overall, the MLPs performance on $T_{all}$ improves over the performance on T1
by 40.1\%. The other methods also improve continuously as more training data is
added. However, the CNN and LSTM improve by only 17\% and 29.4\% with respect to
T1, respectively. Still, this is enough to surpass their full-texts counterparts by
10.1\% and 28.4\%, respectively.

When using as many titles as full-texts, Base-MLP outperforms our proposed
methods clearly by 0.027 points in $F_1$-score. However, Base-MLP does not
benefit as much from additional training data. Its overall improvement is only
20.7\%, so when all titles are used for training, it is outperformed by MLP by a
margin of 5.9\%. On the full-text, MLP has an advantage of 3.6\% over the
baseline. Yet, the baseline still has a large advantage over LSTM and CNN, both on titles and
full-texts.

\paragraph{PubMed}
On PubMed, MLP shows the best full-text performance, whereas LSTM yields the
best results on titles. However, even when all titles are used for training, that is almost 20 times
as many titles as full-texts, the LSTM still lacks behind
the full-text MLP by 2.9\%. However, this is a considerably smaller gap than
when the same number of samples are used. Base-MLP, the best performing method
on \ptone, achieves 10.7\% lower scores than the best method on \pf, which is
MLP.

The MLP and the LSTM show very similar behavior when more training data is
added. This can be seen from Figure~\ref{fig:results-plot} (b), where the lines
of the MLP and LSTM are almost parallel. However, the overall improvement from
T1 to $T_{all}$ is slightly higher for the MLP than for the LSTM. The former
improves by 12.3\%, whereas the latter improves by 10.8\%. The CNN does not seem
to benefit from additional data at all. Initially, on T1, the CNN performs
relatively close to the other methods, lacking behind the LSTM by 5.7\%.
However, as more data is added for training, the CNN demonstrates a worse
classification performance than with fewer training samples. Only when all
available titles are used, the CNN barely outperforms itself on T1 by a very
slim margin of 0.5\%. Consequently, the CNN has the largest difference
to its full-text counterpart out of all proposed methods. It scores 9.8\% lower
on $T_{all}$ than on \pf, whereas the gap is 5.2\% for the MLP and 1.8\% for the
LSTM.

By a considerable margin, the baseline is the best method on T1, where
relatively few samples are used for training. However, despite using 20 times as
many training samples, the performance on $T_{all}$ is only 1.3\% better,
which is a difference to the MLP and LSTM of 3.9\% and 6.2\%, respectively. As
it is the case on the \econ~ dataset, the full-text performance of BaseMLP is 
the second best and gets as close to the MLP as 0.8\%.

\section{Discussion}\label{sec:main-paper:discussion}

The main question of our study is to which extent title-based methods can
catch up to the performance of full-text-based methods by increasing the
amount of title training data. On \econ, the best title-based method outperforms
the best full-text method by 9.4\% when all title training data is
used. Considering that the difference is 16.9\% in favor of the full-text when
the sample sizes are equal, this is an impressive improvement. On PubMed, the
improvement is less astounding. However, the title-based method is close to
competitive to the full-texts, as the difference in score (less than 3\%) is
small. Considering that the gap is much larger for equal sample sizes
(10.7\%), we must acknowledge that current machine learning techniques in
combination with large quantities of data are able to obtain just as good
classification performance by merely using the titles.

However, in order to utilize title-based methods in a
particular application, it is important to understand why there is such a large
difference between the \econ~ and PubMed datasets regarding the benefit of
employing title-based methods with large amounts of data. A possible explanation
for that difference lies in the absolute numbers of full-texts available for
training. As we have pointed out, previous literature suggests that deep
learning models require around 650,000 samples to outperform more traditional
approaches. On \econ, this number of full-texts is far from being reached.
Due to this lack of enough training data, our deep learning models may not
do so well with full-texts, in absolute numbers. The models based on titles on
the other hand may be able to achieve their impressive results because there is
just enough data to unleash the power of deep learning models. In fact, our
MLP, which was optimized for large sample sizes, starts to outperform the
baseline when eight times as many titles as full-texts are used, which nets to
approximately 560,000 training samples. On the PubMed dataset, there are almost
650,000 full-text samples available. Here, the deep learning models can already
work well on full-text. This may explain why the LSTM performs so much better on PubMed's
full-texts than on \econ's full-texts. These findings support the claim from
previous literature that deep learning models work well for text
classification only when the sample size is several hundred thousands.
Furthermore, since our datasets have large label spaces, we can state that this
observation extents to XMLC, which is arguably a
harder task than single-label classification.

In order to push the limits of text classification based on titles, our strategy
was to develop and employ methods that can make use of the vast amount of data
available for training. Our results suggest that this strategy was largely
successful. On both datasets, some of our methods surpass the
performance of the baseline as they are given more and more data for training.
BaseMLP on the other hand cannot make such good use of the additional
training data. This becomes particularly apparent on the PubMed dataset, where
it improves by only 1.3\% even when it has 20 times more training data. Our
methods on the other hand improve considerably the more data is used for
training. To be fair, it is clear that part of the much larger gain compared to
the baseline is due to overfitting on the small title datasets such as T1.
Recall that our methods are optimized towards their performance on $T_{all}$. This design
decision in our study was made to be able to fairly assess the development of
the performance as the sample size increases.
We observed that the capacity of the resulting models is likely too large for smaller
datasets, which results in overfitting.
This can be seen by the fact that
BaseMLP, which has considerably lower capacity, outperforms our proposed methods
on both PubMed and \econ. Yet, on $T_{all}$, MLP outperforms Base-MLP by a
wide margin on both datasets. LSTM is close to Base-MLP on \econ~ and
outperforms it drastically on PubMed. Again, this indicates the success of our
strategy. The only exception is the CNN, which does not benefit as much from
additional data as our other proposed methods, although its capacity is large
compared to other architectures recently proposed in the literature.
It is particularly interesting that the performance of CNN (and Base-MLP, too)
actually has a drop on PubMed as more title samples are used for training. We
explain this by the fact that in T2 to T8 more than 1,300 new labels that do
not occur in T1 are introduced. Hence, the models have to account for
these new labels even though they never appear in the test set, reducing their
capacity to learn to classify the labels relevant to the test set.

Considering the amount of attention CNNs have received in recent
years for their performance in text classification (cf.
Section~\ref{sec:main-paper:relevant-work}), the results of our CNN is rather
underwhelming. This is true for both full-text, and titles.
In neither case is this due to overfitting. Our preliminary experiments showed
that the CNN benefits from increasing the feature map size a lot. Yet, the CNNs
do not benefit from additional training data in the same way LSTMs and MLPs do.
On PubMed, no benefit at all can be observed. On \econ, the rate of
improvement is comparable to MLP and LSTM only up to a factor of four times the
number of full-texts. After that, the improvement is relatively
marginal. In conclusion of our findings, we think it would be good if future
research shifted its focus more towards MLPs and LSTMs, as they have
demonstrated to be serious competitors for CNNs in text classification.

The goal of this study is to investigate to which extent a model trained on vast amounts of
sample titles can compensate for the lack of information in comparison to the
full-text. 
Thus, it is not the aim to achieve results beyond the state-of-the-art
performance on, e.\,g., full-texts.
Yet, the models we use are based on and
also enhance recently proposed models. As described in
Section~\ref{sec:main-paper:training-procedure}, the MLP is at its core a
non-linear version of the popular fastText. The CNN is based on recent advances
from the domain of text-based XMLC, but was improved by
integrating more fine-grained window sizes and larger feature maps. Finally, we
present a strong bidirectional LSTM with attention over the outputs that does
not assume a hierarchical structure of the document and therefore also works
for short text snippets, in contrast to previous work by Yang et
al.~\cite{yang2016hierarchical}. Therefore, our methods are good candidates for
researchers to also adopt for single-label text classification.

A common problem in machine learning research and in deep learning in
particular is that models are very sensitive to the choice of hyperparameters. However,
examining the whole hyperparameter space is very difficult due to its
combinatorial complexity. Recently, this has called the validity of
deep learning results into question, for example in
language-modeling~\cite{melis2017state} or even text
classification~\cite{le2017convolutional}. This problem persists in our study
as well. However, instead of simply assuming values for our parameters or
manually tuning them in a somewhat arbitrary fashion, we took an incremental
tuning approach that in the end led to an improvement over initial base models
from the literature in all cases. This gives us reason to believe that our
results are largely reliable.

In this study, we have compared three deep learning methods. 
We did not compare against linear models such as logistic regression, and we did
not compare against other non-linear approaches such as kNN or
SVMs. However, as described in
Section~\ref{sec:main-paper:relevant-work}, there is evidence that traditional
linear methods are inferior to non-linear ones when the training data is large.
More importantly, we compared against a non-linear baseline by Galke et
al.~\cite{galke2017multilabel} that was shown to outperform not only linear
models, but also other non-linear, non-parametric models like kNN and SVMs on
a diverse set of datasets and by a wide margin on both titles and full-texts.

For comparability, our proposed models were chosen such that they can be
employed to titles and full-texts uniformly. This prohibits that certain
strengts of the full-text can be made full use of.
For instance, full-text models might benefit greatly from a hierarchical model
as proposed by Yang et al.~\cite{yang2016hierarchical}. On the other hand, we
tried our best to tap the full potential of full-texts. For instance, in our CNN
we employ dynamic max-pooling with $p = 3$ for full-texts although this
does not have any beneficial effect on titles.

In our study, we have examined two datasets from digital libraries of
scientific content. We argue that these results are likely to generalize to
other datasets of scientific publications as well. In consent with previous
text classification research, we found our deep learning methods to require approximately 550,000
samples to outperform the previous baseline (\eteight). While this number of
titles can certainly be reached in domains other than economics and
biomedicine, not many scientific domains will reach this number of full-texts.
This can be seen by considering the fact that the availability of full-texts is
generally tied to their open access rate. The rate of open access journals in
academia is approximately 7\% as reported by Teplitskiy et
al.~\cite{teplitskiy2017amplifying}. This number closely matches the rate of 5
to 6.5\% of available full-texts in our datasets. Moreover, Teplitskiy et al.'s
study also suggests that the corpus of publications from the medical domain are
among the largest.
Therefore, we think it is likely that in other domains at least the relatively
small gap of less than 3\% between titles and full-texts can be achieved as
well.
However, in many other domains where only few full-texts are available, training models
on titles may actually be much better, as we demonstrated for the economics
domain.

Our results are of great practical importance for automatic semantic indexing
in digital libraries. Considering the large amount of new literature published
every year, subject indexers rely on the assistance of machines.
It is desirable to have algorithms that produce good annotation suggestions by
using only the title as textual input instead of the abstract or full-text. This
is because the title is easy to obtain and free to use in text mining
applications, whereas the full-text and even the abstract are often either not
available or may not be processed automatically due to legal restrictions.
For instance, in EconBiz only approximately 15\% of the documents have an
abstract and 7\% have a full-text available.
Furthermore, the task of downloading and processing the full-text is
cumbersome, where one has tens of gigabytes of data. In contrast, as
Table~\ref{tab:datasets} shows, titles only comprise up to a few gigabytes of
data.
Thus, the full-text are by an order of magnitude larger. Thanks to
the vast amount of data available for training title-based deep learning
models, our study demonstrates that in a realistic scenario deep learning
algorithms are able to satisfy the demand for sufficiently strong
title-based classification methods.
Engineers of (semi-)automatic semantic indexing algorithms should therefore
consider shifting their focus from full-text-based classification to
title-based classification in order to maximize the applicability of automatic
semantic indexing systems.
\section{Conclusion}
In this paper, we have successfully answered the question if a semantic indexing
system based on the title can reach the performance of a system based on the full-text if the number of
samples for training the title-based method is much larger than the number of
full-text samples. To this end, we developed three deep learning methods and
evaluated them on two scientific datasets of different size. We found that
in one case such a system is competitive with the full-text system, and in the
other case it even yields considerably better scores. These results have
important implications for automatic semantic indexing systems in digital
libraries of scientific content.

In the future, we want to push the limits of classification based on large
amounts of titles even further, and we would like to encourage the community to
do the same.
To this end, we published the title datasets, the source code, and an extended
version of this paper on
GitHub\footnote{\url{https://github.com/florianmai/Quadflor}}.

\begin{acks}
This research was co-financed by the \grantsponsor{eu-sponsor}{EU H2020}{}
project MOVING (\url{http://www.moving-project.eu/}) under contract no
\grantnum{eu-sponsor}{693092}.
We would like to thank Tamara Pianos and Tobias Rebholz from ZBW for providing
helpful information on the EconBiz dataset.
\end{acks}

\bibliographystyle{ACM-Reference-Format}

\bibliography{paper-short}

\end{document}